\documentstyle[epsfig,mathptm]{mn} 
\newif\ifAMStwofonts
\AMStwofontstrue                % <- comment this out if AMS fonts not available
\DeclareMathVersion{bold}                    % <- seems to be needed with mathptm
\DeclareMathAlphabet{\mathbfit}{OT1}{cmr}{bx}{it}
\SetMathAlphabet\mathbfit{bold}{OT1}{cmr}{bx}{it}
\DeclareMathAlphabet{\mathbfss}{OT1}{cmss}{bx}{n}
\SetMathAlphabet\mathbfss{bold}{OT1}{cmss}{bx}{n}
\ifAMStwofonts
  \ifCUPmtlplainloaded \else
    \DeclareSymbolFont{UPM}{U}{eur}{m}{n}
    \SetSymbolFont{UPM}{bold}{U}{eur}{b}{n}
    \DeclareSymbolFont{AMSa}{U}{msa}{m}{n}
    \DeclareMathSymbol{\upi}{0}{UPM}{"19}
    \DeclareMathSymbol{\umu}{0}{UPM}{"16}
    \DeclareMathSymbol{\upartial}{0}{UPM}{"40}
    \DeclareMathSymbol{\leqslant}{3}{AMSa}{"36}
    \DeclareMathSymbol{\geqslant}{3}{AMSa}{"3E}

    \let\leq=\leqslant 

  \fi
\fi

\newcommand{\mnras}{{MNRAS}}

\newcommand{\aap}{{ A\&A}}

\newcommand{\apj}{{ ApJ}}
\def\kms{\ifmmode {\rm \ km \ s^{-1}}\else$\rm km s^{-1}$\fi}

\def\s8{{\sigma_8}}

\title
[MEM lens reconstruction using shear and/or magnification data]
{Maximum-entropy reconstruction of gravitational lenses using
shear and/or magnification data.}
\author[S.L.~Bridle et al.]
{S.L.~Bridle$^{1,2}$, M.P.~Hobson$^2$, Richard~Saunders$^2$,
A.N.~Lasenby$^2$\\ 
$^1$ 
Laboratoire d'Astrophysique, 
Observatoire Midi-Pyr\'en\'ees, 
14 Avenue E. Belin, 31400 Toulouse, France\\
$^2$ Astrophysics Group, Cavendish Laboratory,  Madingley Road, 
Cambridge CB3 0HE, UK\\
}

\date{Accepted ???. Received ???; in original form \today}
\pagerange{\pageref{firstpage}--\pageref{lastpage}}
\pubyear{1999}
\begin{document}
\maketitle
\label{firstpage}
\begin{abstract} 

We demonstrate that the maximum-entropy method for gravitational lens 
reconstruction presented in Bridle et al. (1998) may be applied even when only 
shear \emph{or} magnification information is present. We also demonstrate that the 
method can easily handle irregularly shaped observing fields and, because shear is 
a non-local function of the lensing mass, reconstructions that use shear 
information can successfully bridge small gaps in observations. 
For our simulations we use a mass density distribution that is realistic for a 
$z=0.4$ cluster of total mass around $10^{15} h^{-1} M_{\odot}$. Using HST-quality 
shear data alone, covering the area of four WFPC2 observations, we detect $60$ per 
cent of the mass of the cluster within the area observed, despite the mass sheet
degeneracy. This is qualitatively because the shear provides information about the 
variations in the mass distribution, and our prior includes a positivity 
constraint. We investigate the effect of using various sizes of observing field and
find that $50$ to $100$ per cent of the cluster mass is detected, depending on the 
observing strategy and cluster shape. Finally we demonstrate how this method can 
cope with strong lensing regions of a mass distribution.

\end{abstract}
 
\section{Introduction}
\label{intro}

The study of gravitational lensing by clusters of galaxies is
important for cosmology as a probe of both dark and
luminous matter. `Weak lensing', the very slight shearing and magnification of
background galaxies is much more prevalent than
the more spectacular `strong lensing' effects, such as multiple or highly
distorted images.
See Mellier (1999) and Bartelman \& Schneider (2000) for recent reviews.
Many methods have now been proposed for reconstructing the 
cluster mass distribution from weak lensing observations and many 
authors have applied these methods to observations in
order to reconstruct the 
mass distributions. The problem
splits naturally into two stages. Firstly the problem of
estimating a shear and/or magnification maps from observations, and
secondly estimating the 
mass density distribution from the
shear and/or magnification maps. In this paper we are concerned with the
second problem.

The first non-parametric method for inverting shear data to estimate
the mass distribution was proposed by Kaiser \& Squires (1993)
\& used just the shearing of background galaxies. This method has
now been improved by several authors to circumvent many of its 
original problems (Schneider \& Seitz 1995; Seitz \& Schneider 1995; Kaiser
1995; Squires \& Kaiser 1996). However, Kaiser \&
Squires, Schneider \& Seitz and others have indicated
that even perfect shear information is insufficient to reconstruct the
projected mass unambiguously due to the `mass sheet
degeneracy'. More recent methods have therefore combined shear
information with magnification information in order to overcome this
problem (Bartelmann et al 1996; Seitz, Schneider \& Bartelmann 1998,
Bridle et al. 1998). Because these methods iteratively find the mass
or gravitational 
potential distribution which best fit the observations, they also
can easily overcome the difficulties of the original Kaiser \&
Squires method.

It has also been noted that magnification data alone are
enough to reconstruct the lens (Broadhurst, Taylor \& Peacock 1995;
Dye \& Taylor 1998). Besides having the potential to avoid the mass sheet
degeneracy problem, an advantage of using magnification data alone is
that the point spread function of the telescope and the circularising
effects of the atmosphere are much less of a problem (Broadhurst et
al. 1995). It does however require the sizes and/or number densities of
\emph{unlensed} background galaxies to be known. In addition,
one has to know -- or assume -- the luminosity,
redshift and spatial distribution of background galaxies (e.g. Dye \& Taylor
1998).  

In an earlier paper (Bridle et al. 1998, hereafter BHLS) we presented
a maximum-entropy method for reconstructing the projected mass
distribution from both shear and magnification data. 
In contrast to comparable
methods, we reconstruct the projected mass density distribution itself,
rather than reconstructing the two-dimensional Newtonian potential and
then converting this into the mass distribution. This seems to us to
be a more elegant and only slightly more computationally expensive
method. As a result, we are then able to use the shape of the
probability function in order to estimate the errors on our
reconstruction. In addition, the method allows reconstruction outside the
observed field, which makes fuller use of the shear information.
In this paper we
demonstrate the way in which this method may be extended to
the cases where no shear information is available, or where no
magnification data are available, and illustrate the effect on the
resulting reconstructions. 
An evaluation of the relative merits of using shear versus magnification
information, along with a comparison of reconstructions for a two 
parameter lens model, is given in Schneider, King \& Erben (2000).
We highlight the flexibility of our method in analysing
observations of irregularly-shaped patches of sky and reconstructing
mass distributions containing strong lensing regions.

\section{Method}
\label{method}

We follow the method described in BHLS, in which the most
probable mass distribution is inferred from ellipticity and/or
magnification data (each with estimated errors) under the assumption of an
entropic prior on the mass distribution with a uniform default model.
This method finds the most probable mass density distribution given the data and
the entropic prior. The probability of a given mass
distribution is $\propto \exp{(-\chi^2+\alpha S)}$ where $\chi^2$ is from
the difference between the observations and the predictions and
$\alpha S$ is the maximum entropy prior. 
As in BHLS, we use a simple simulation method to produce simulated
data from a lensing mass distribution. 

\begin{figure}
\centerline{\vbox{
\epsfig{file=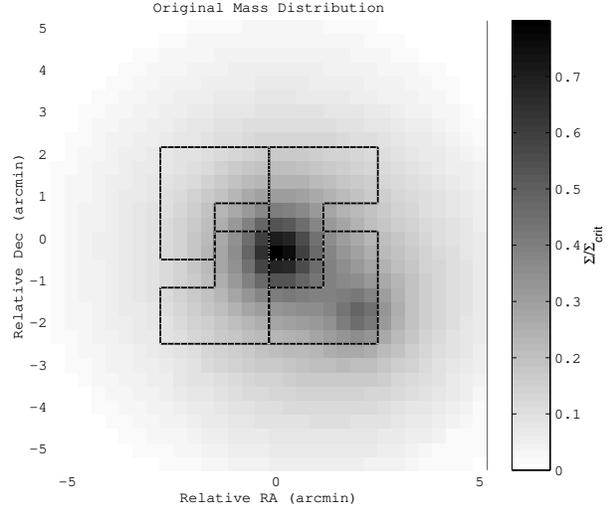,height=8cm, angle=90}
}}
\caption{
Original mass distribution used for the simulations. 
The dashed lines contain the regions of sky for which simulated
observations are made.
\label{origmass}}
\end{figure}

The lensing mass distribution
we use for the simulations is plotted in Fig. \ref{origmass} and
described in Section \ref{mass}.
The ellipticities and magnifications expected
from this mass distribution are calculated at each point on the same
grid of pixels used for the mass distribution. For convenience we
consider as input data the inverse magnification rather than the
magnification. 

We then simulate observations from a finite patch of sky.
In BHLS we
simulated an observation of a square patch of sky, yet one great
advantage of our $\chi^2$ type of method is that the observed patch
of sky can be of any shape. Therefore in this paper we take the
opportunity to demonstrate this, and the patch of sky observed is made 
up of four HST WFPC2 pointings which do not fit together perfectly,
shown by the dashed lines in Fig. \ref{origmass}. 

We then add random noise to each shear data point to simulate the effect of
the intrinsic galaxy ellipticities.
In this paper, as in BHLS, we add Gaussian noise of mean zero
and standard deviation $0.05$ to the ellipticity data points, 
Within the observed regions enclosed by dotted lines in Fig. \ref{origmass} ,
the peak ellipticity from this mass density distribution is $0.4$ and
the mean is $0.19$. This therefore 
represents a signal-to-noise ratio of at most $8$, and on average
$4$. 
The average signal-to-noise ratio is about that inferred from shear
measurements based on HST imaging of A2218 (Smail et al. 1997),
which are typical of those achievable with the HST on massive clusters
at $0.1<z<0.5$. 
This is also similar to the theoretical noise expected from
random galaxy ellipticities from $\approx 20$ background galaxies, as
calculated by Schneider \& Seitz (1995). 

We add Gaussian noise of mean zero and standard deviation $0.1$ to the
inverse magnification data.
The inverse magnification, $r$ ranges
from $0.02$ to $0.99$ with $\left< 1-r \right>=0.47$ (note that $r=1$ for
no lensing). Thus
the signal-to-noise ratio peaks at $(1-0.04)/0.1=9.8$ and is on average
$5$. For $25$ galaxies per pixel, Poisson noise gives a
signal-to-noise ratio of $5$. 

We do not include systematic errors in our simulations. 
For accurate shear estimation it is necessary to correct for
additional distortions due to the atmosphere and telescope, 
but much effort has gone into understanding
how to correct for this using stars in the 
image (Kaiser, Squires \& Broadhurst 1995).
In addition, the ellipticity distribution of the particular 
background galaxies will be similar to that in a random patch of 
sky and is therefore reasonably well known.
To estimate magnifications the number densities or sizes of
galaxies are compared to those in an unlensed field.
However, the apparent optical fluxes, sizes and number counts of
galaxies behind a particular cluster may 
be affected by clumping and extinction, so that systematic 
errors may be of greater consequence than random ones.

\subsection{The Mass Distribution}
\label{mass}

The mass density distribution we use for our simulations is 
shown by the greyscale in Fig. \ref{origmass}
and consists of two clumps each with a density profile
based on a King model.
Because real clusters are finite in extent, we do not use a King
profile as our mass distribution.
Instead we truncate the 3d King profile at some cut
off radius. This also conveniently avoids the problem that strict King
models have a total mass that does not converge. 
The mass density as a function of position ${\bf r}$ relative to the cluster
centre is
\begin{equation}
\rho({\bf r}) = 
\left\{
\begin{array}{ll}
\frac{\rho_{0}}{1+\left({|{\bf r}|}/{r_{\rm c}} \right)^2} &
\qquad r \leq r_{\rm f}\\
0& \qquad r > r_{\rm f},
\end{array}
\right.
\end{equation}
where $r_{\rm f}$ is the cutoff radius and $r_{\rm c}$ is the core radius.
Projected onto the plane of the sky and converted to angular postions
($\theta=r/D_{\rm d}$, where $D_{\rm d}$ is the angular diameter distance to the lens)
this can be shown to be 
\begin{equation}
\Sigma{\left(\bf{\theta}\right)}
=
\frac{2 \Sigma_0 \theta_c^2}{\sqrt{\theta_c^2+\bf{\theta}^2}}
\arctan{ \frac{\sqrt{\theta_f^2-\bf{\theta}^2}}{\sqrt{\theta_c^2+\bf{\theta}^2}}}
\end{equation}
where $\Sigma_0=\rho_0 D_{\rm d}$.
The
central cluster the core radius $\theta_{\rm c}=0.6$ arcmin, typical for a
big cluster at redshift $z=0.4$, corresponding to
a proper distance of $120 h^{-1}$ kpc (defining
$h=H_0/(100$~\mbox{$\rm{km} \rm{s}^{-1} \rm{Mpc}^{-1}$}$)$). Both
clusters have $\theta_{\rm f}=10\,\,\theta_{\rm c}$ and
the central cluster alone has a peak of
$\Sigma/\Sigma_{\rm{crit}}=0.7$, where $\Sigma_{\rm{crit}}$ is the
critical density, given by
\begin{equation}
\Sigma_{\rm{crit}}=\frac{c^2}{4\pi G}\frac{D_{\rm{s}}}{D_{\rm{d}} D_{\rm{ds}}}
\label{sigmaeqn}
\end{equation} 
where $D_{\rm{d}}$, $D_{\rm{s}}$ and $D_{\rm{ds}}$ are the angular
diameter distances from the observer to the lens, the observer to the
source, and the source to the lens respectively. Thus for the central
cluster at $z=0.4$ and the background galaxies at $z=\infty$, the
central cluster would have a mass of $9.2 \times 10^{14} h^{-1}
M_{\odot}$. Therefore it
represents a massive, yet not quite critical, cluster. 
With the addition of the smaller clump, 
the total maximum is 
$\Sigma/\Sigma_{\rm crit}=0.79$.
For comparison with the equivalent values found in the
reconstructions, the total mass inside the observing boxes would be
$5.41\times 10^{14} h^{-1} M_{\odot}$, given the lens geometry described above. 

\subsection{Handling missing data}

In BHLS we assumed that both ellipticity and magnification data for a
square patch of sky were
present, but in this paper we investigate the effect of having pieces
of data missing: no magnification data; no shear data; irregularly
shaped observations; and gaps in sky coverage. It is necessary to
explain how missing data are in practice handled by our method.

If a data element is unavailable, then this is equivalent to 
an infinitely large error on this element. Since it is always the
reciprocal of the error (the statistical weight) that is required in
$\chi^2$ and its derivatives then there is no need to handle
infinities in the reconstruction program and missing data are simply
assigned a statistical weight of zero.
The number of data points with which to compare $\chi^2$, for example, 
is then equal to the number of data points which have finite error.
By setting the statistical weights of the ellipticity
information to zero we cause the method to reconstruct from only the
magnification data, and vice versa for ellipticity data alone. 

\subsection{Setting the maximum-entropy prior}
\label{prior}

The maximum-entropy prior requires a `model' to be specified, which
is the reconstruction that would be
obtained in the absence of any evidence in the data to the
contrary. As in BHLS we assume a `flat' model, that is, one having a
constant value across the entire map. For all of the reconstructions
in this paper, we assume a level for our flat model of
$\Sigma/\Sigma_{\rm crit}=0.02$. 
Setting the model to such a low value ensures that there will be
mass peaks at a level significantly greater than $0.02$ only if the data
point to it strongly enough.  

In BHLS we investigated the effect of varying the model level on the
reconstruction from both shear and magnification data. We found that
varying its value by an order of magnitude either way had negligible
effect on the reconstruction, within the observed region. We find the
same result for the simulation in this paper for reconstruction from
magnification data alone, as well as for the reconstruction from both
shear and magnification data. For the reconstruction from shear data
alone we find that reducing the model value makes no difference to the
reconstruction. However, increasing the model value effectively
transforms the reconstructed mass distribution by the mass sheet
degeneracy transformation, to higher values.

A parameter 
of the prior that is left to be determined is the
weighting of the prior relative to the data, 
$\alpha$. 
In BHLS we described a Bayesian technique for fixing the value of
$\alpha$ (following Skilling 1989 and Gull \& Skilling
1990) but in
fact went on to use a quick approximate method which we found to agree to
within a few percent. 
We find the same result for the reconstructions shown in this paper,
although, as expected, if very much less and/or much noiser data is
used then the methods differ.

\section{Results}
\label{results}

\begin{figure*}
\centerline{
\begin{tabular}{cccc}
(a)&
\mbox{\epsfig{file=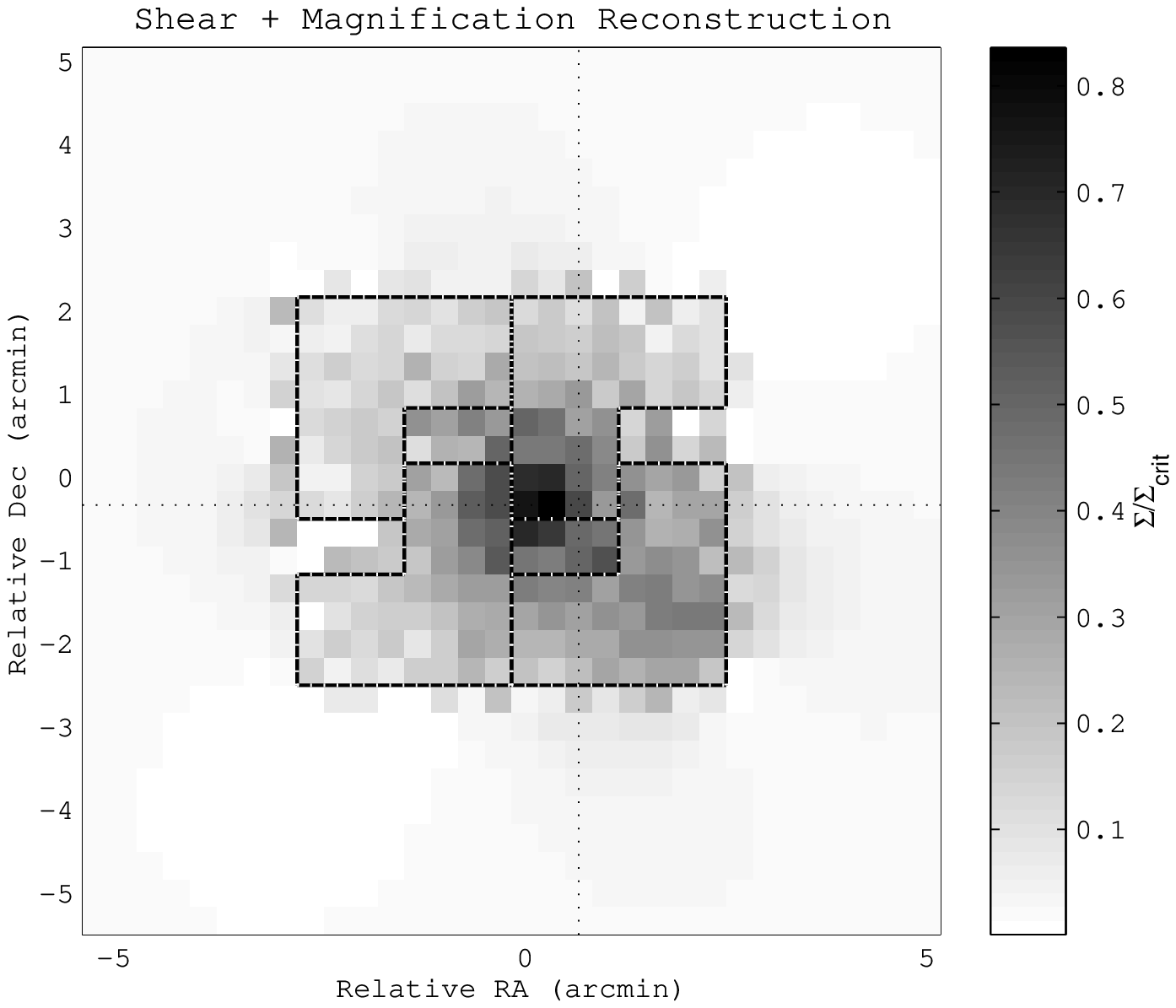,height=8cm, angle=90}}&
(b)&
\mbox{\epsfig{file=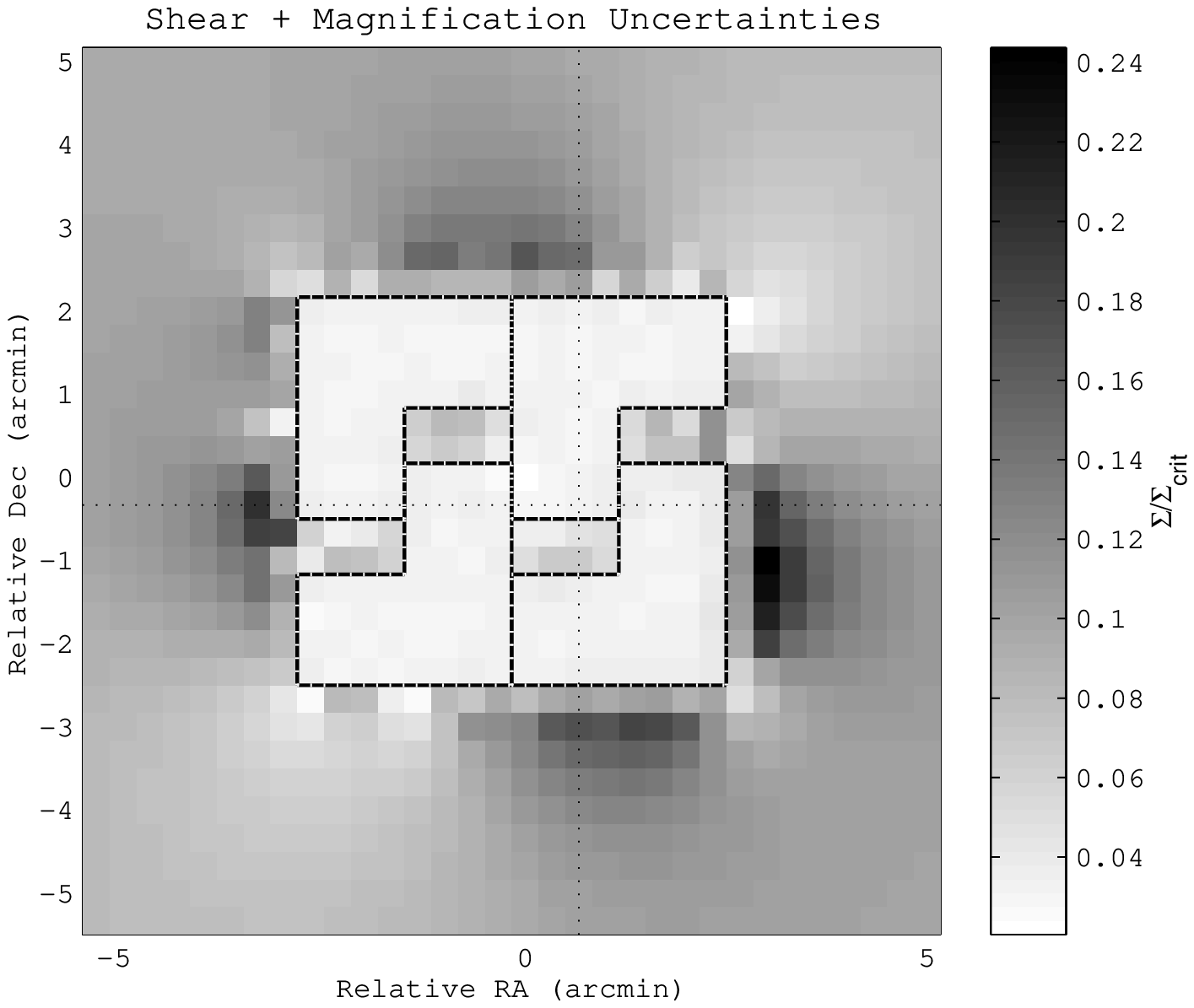,height=8cm, angle=90}}\\
(c)&
\mbox{\epsfig{file=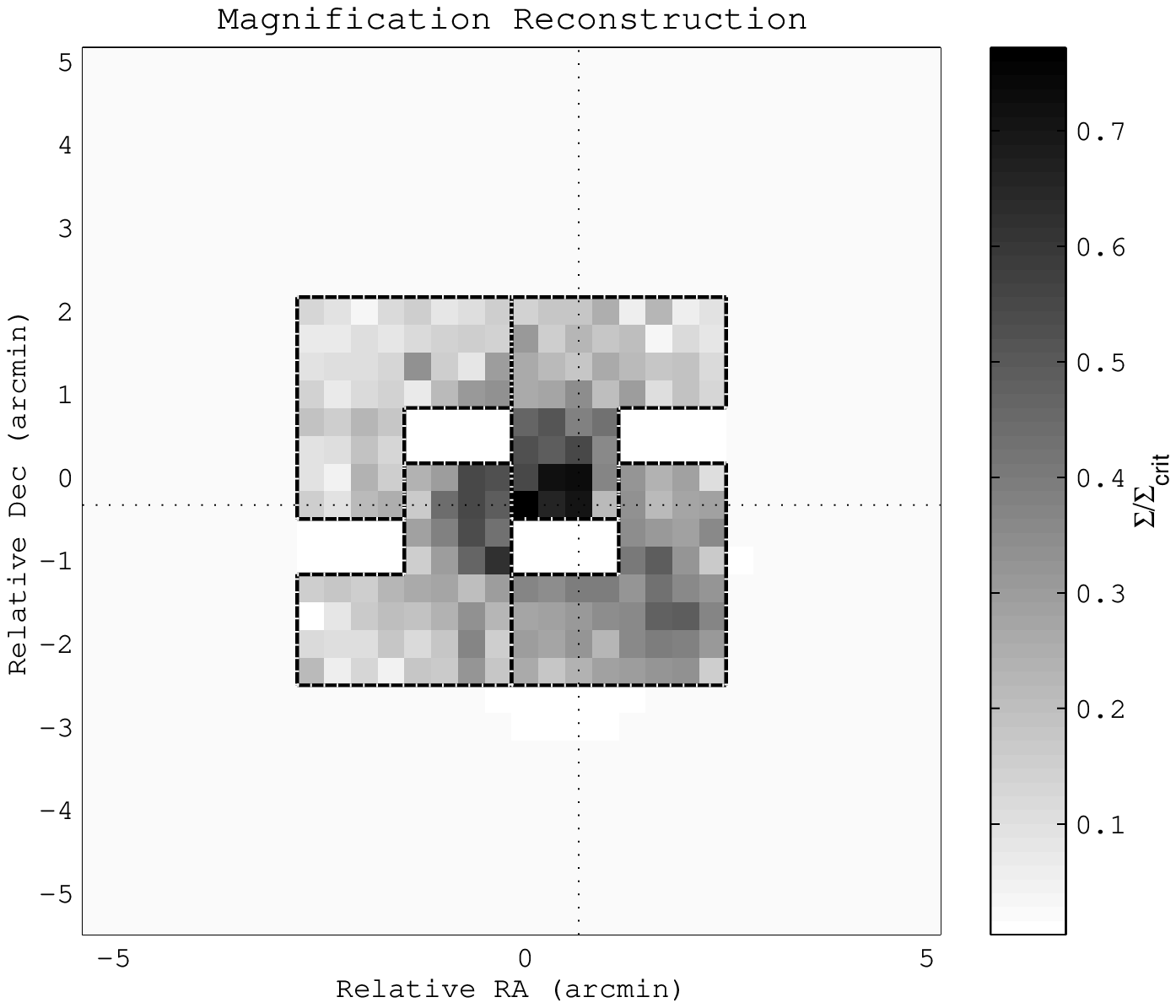,height=8cm, angle=90}}&
(d)&
\mbox{\epsfig{file=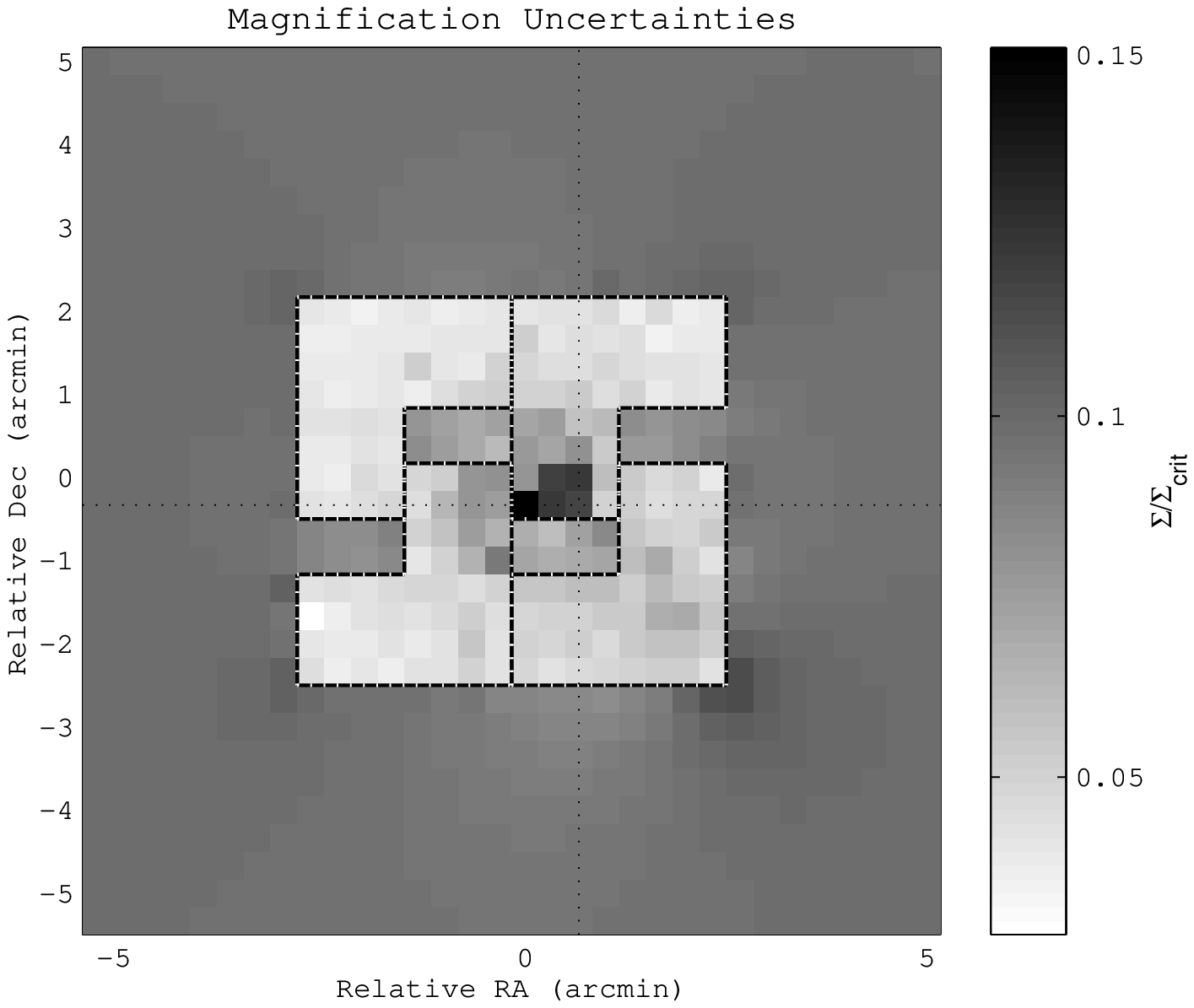,height=8cm, angle=90}}\\
(e)&
\mbox{\epsfig{file=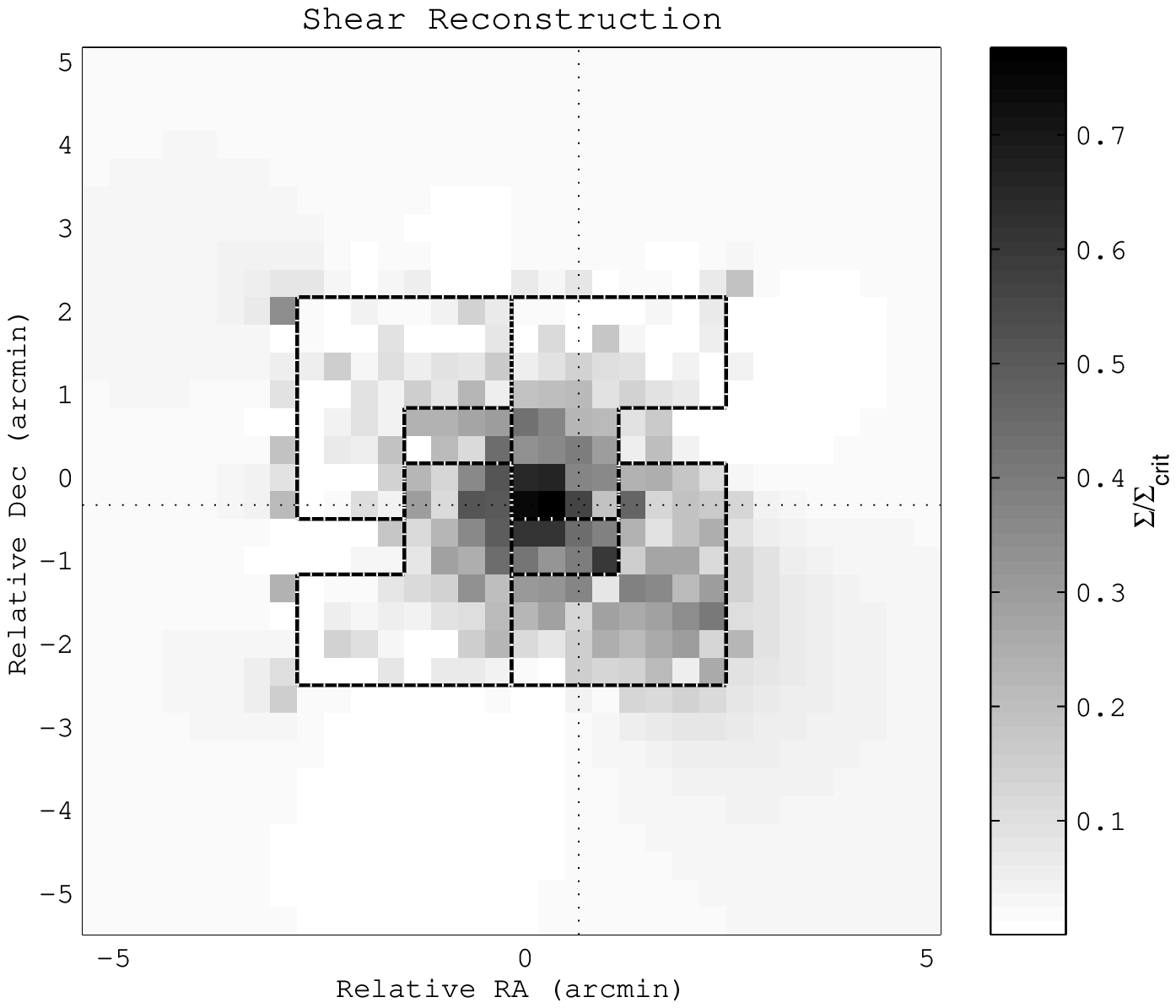,height=8cm, angle=90}}&
(f)&
\mbox{\epsfig{file=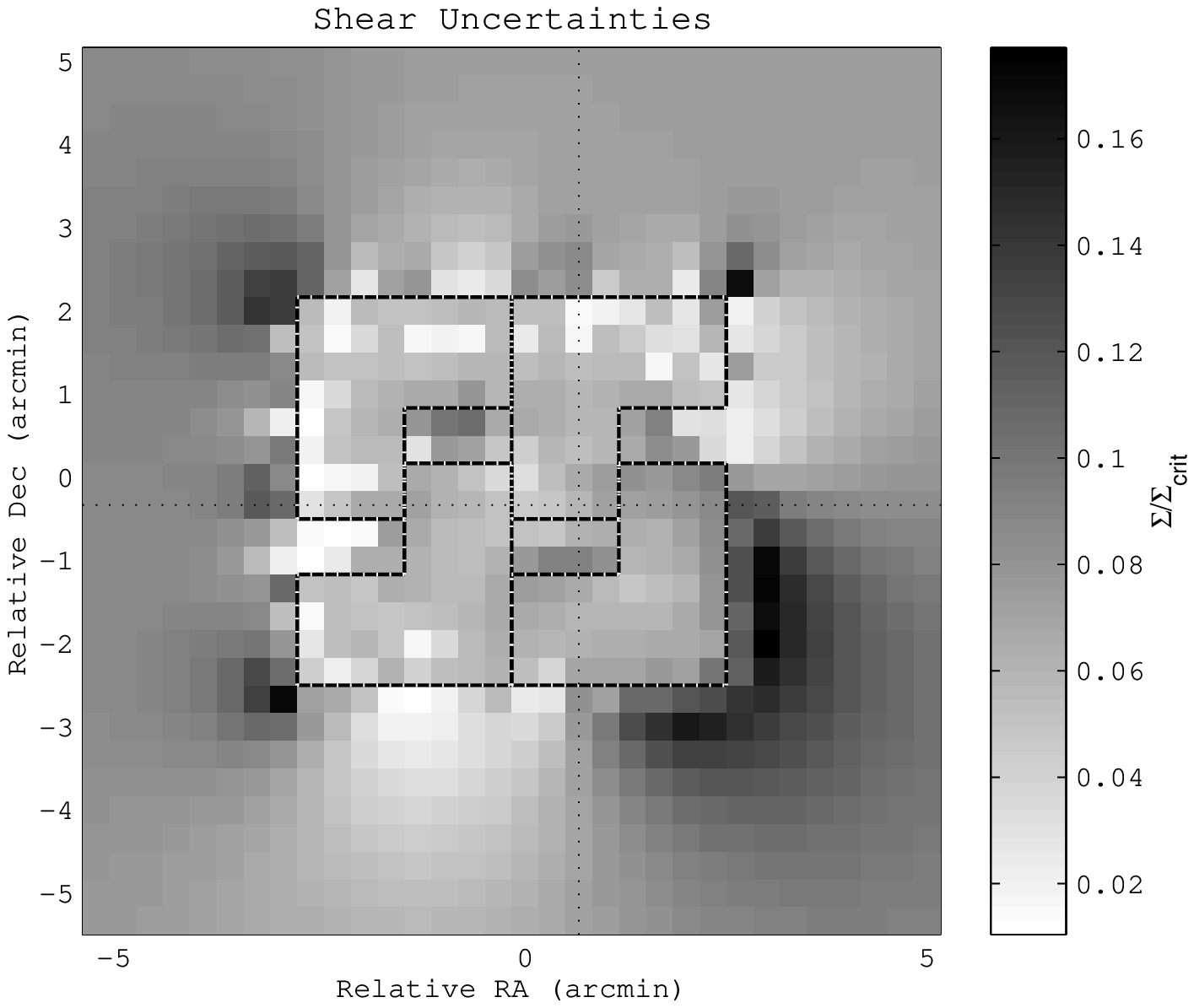,height=8cm, angle=90}}
\end{tabular}
}
\caption{
(a) Mass distribution reconstructed using shear and magnification
information. (b)
Errors on the mass distribution reconstructed using shear and
magnification information. The dotted lines show the lines along which
the cross sections are taken.
(c) Mass distribution reconstructed using only magnification information.
(d) Errors on the mass distribution reconstructed using only
magnification information. The dotted lines show the lines along which
the cross sections are taken.
(e) Mass distribution reconstructed using only shear information.
(f) Errors on the mass distribution reconstructed using only shear information.
The dotted lines show the lines along which
the cross sections are taken.
\label{images4}}
\end{figure*}

\subsection{Both shear and magnification data}

The result of reconstructing from both shear and magnification data is
shown in Fig. \ref{images4} (a). As detailed in BHLS, we estimate the
marginalised errors on each pixel using the curvature matrix of the logarithm
of the probability function evaluated at the best-fit point.
These are plotted in Fig. \ref{images4} (b).
This reconstruction demonstrates clearly the ability of the method to
cope with irregularly-shaped data fields. 
As will be demonstrated more clearly below, since \emph{shear} data
provide non-local information about the mass distribution it is
possible for the reconstruction to  bridge (small) gaps between the
observed patches of sky. Also some of the mass of the smaller clump,
which is outside the observed field, is detected.

Assuming the lens geometry described in Section \ref{mass}, the total
mass in the observing area in the reconstruction is $5.15\times
10^{14} h^{-1} M_{\odot}$, and the rms mass within the observing area on
the error estimates map is $0.16\times 10^{14} h^{-1} M_{\odot}$, 
compared to $5.41\times10^{14} h^{-1} M_{\odot}$ for the original mass distribution. 
Thus not only
has the shape of the mass distribution been well reproduced, but also
the total mass in the observed field is well reproduced and the error
estimate gives a good indication of the uncertainty in this quantity. 

\begin{figure*}
\centerline{
\mbox{
\epsfig{file=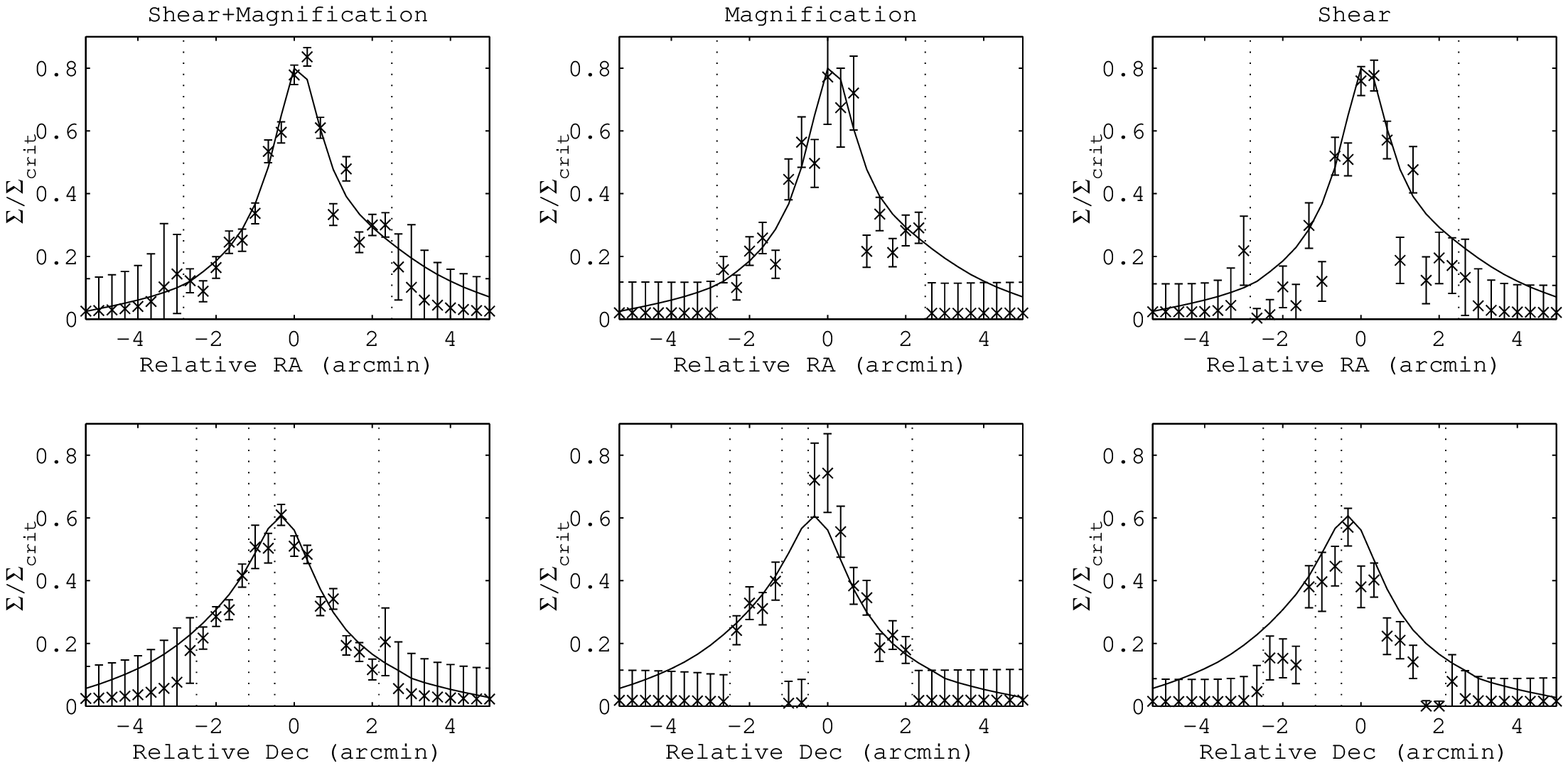,height=17cm, angle=90}
}
}
\caption{
Cross sections through the mass reconstructions shown in Fig.
\ref{images4}. Solid lines
are the original mass distribution, crosses are the reconstructed mass
distribution, and the error bars are those found from the curvature 
matrix.
 The edges of the
observed field are shown by the dotted lines.
 Upper
panels show slices at a constant Dec of $-20$ arcsec (halfway between
the gaps in the observations) for
reconstructions from shear+magnification, magnification and shear
respectively. Lower panels show slices at a constant RA of $40$
arcsec, the gap in the observations is clearly seen in the
magnification reconstruction.
\label{cross}}
\end{figure*}

Two cross sections through the reconstruction are shown in the left
hand column of Fig. \ref{cross}. The original mass distribution is
shown by the solid line and the reconstructed values by the
crosses. The error bars are those found from the curvature matrix.
The edges of the observation field are the dotted lines.
The errors can be seen to be a reasonable indication of the
differences between the original and reconstructed mass distributions
within the observed area and a short distance beyond.

\subsection{Magnification data alone}

The result of reconstructing from magnification data alone 
is shown in Fig. \ref{images4} (c), with errors as shown in
Fig. \ref{images4} (d).
This clearly illustrates the fact that the magnification data
gives very little information about the mass outside the observed
field. 
The cross sections plotted in
the middle panel of Fig. \ref{cross} show that the reconstruction
inside the observed field is reasonable.  
In addition, the total mass in the observing box, $5.15\times 10^{14}
h^{-1} M_{\odot}$ plus or minus $0.25\times 10^{14} h^{-1} M_{\odot}$ may be compared
with that of the true mass distribution, and seen to be in agreement.

\subsection{Shear data alone}

The result of reconstructing from shear data alone is shown in
Fig. \ref{images4} (e), with errors shown in Fig. \ref{images4}
(f). 
The most striking point is that, at first sight,
this reconstruction is not very
much worse than that from using both shear \emph{and} magnification data
(Fig. \ref{images4} (a)). On closer inspection it is clear
that this reconstruction contains less mass and is a little more noisy. 
The total mass inside the observed field in this reconstruction is 
$3.22\times 10^{14} h^{-1} M_{\odot}$ and the sum of
the rms errors in this field $0.28\times 10^{14} h^{-1} M_{\odot}$. This
compares with $5.41\times 10^{14} h^{-1} M_{\odot}$ for the original mass
distribution. So despite the mass sheet degeneracy, $60 \pm 5$ 
per cent of the
mass has been detected. Qualitatively this is because the shears
provide information about the spatial variations in the mass, and the
prior constrains the mass to be positive. It is also significant that
because we take into account the errors on the shears, and regularize
with a prior, we do not over fit to the noise, and only find mass where
there is sufficient evidence in the data. 

Note that in places where the reconstructed mass is very close to
zero, the errors tend to be unrealistically small. In the calculation of
the errors on the mass distribution we have assumed that the
(posterior) probability distribution is Gaussian about the best fit
point. However, this is unlikely to be the case in places where the
mass distribution is close to zero, hence the small error estimates.

\subsection{Other observing strategies}

\begin{figure*}
\centerline{
\begin{tabular}{cccc}
(a)&
\mbox{\epsfig{file=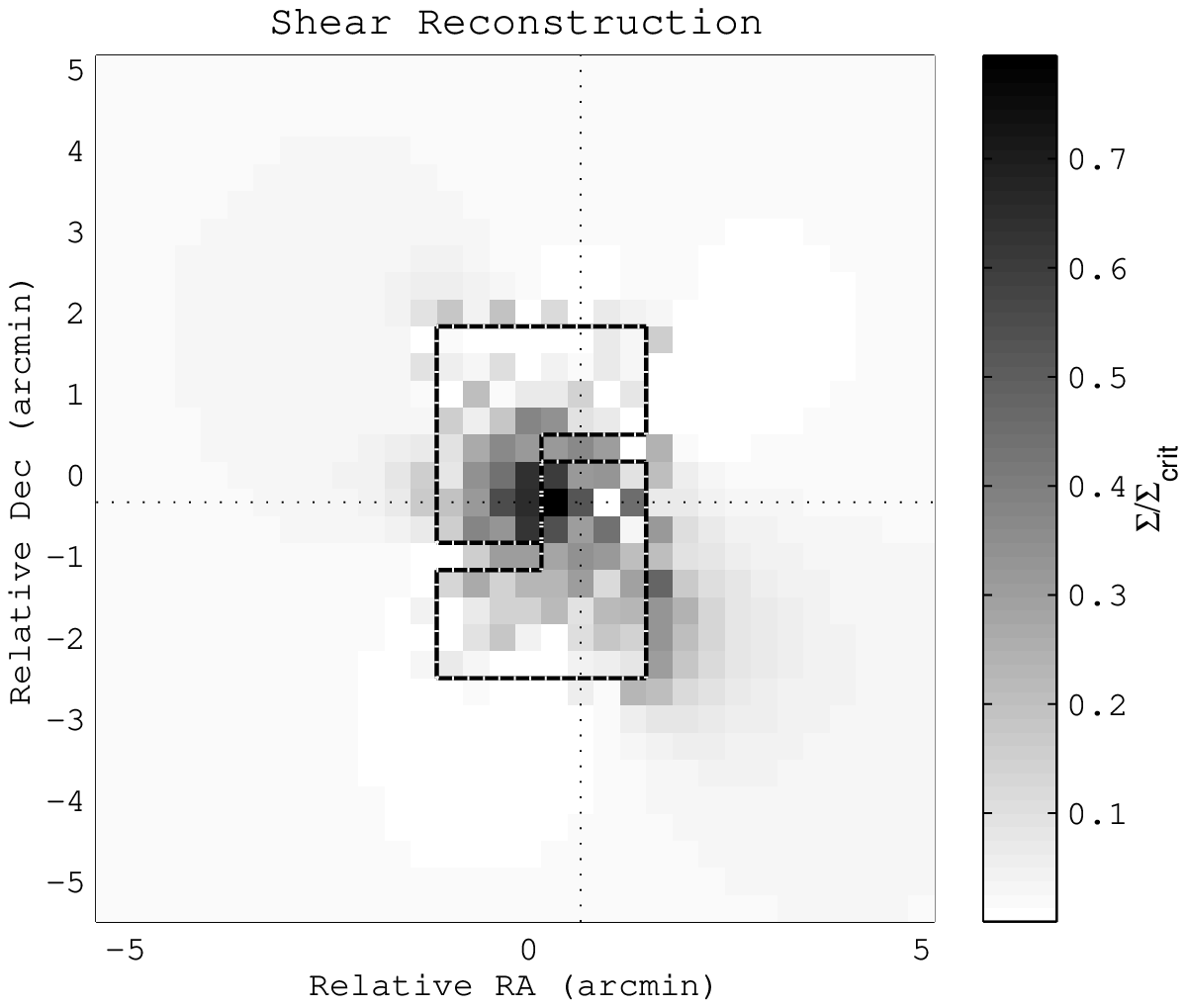,height=8cm, angle=90}}&
(b)&
\mbox{\epsfig{file=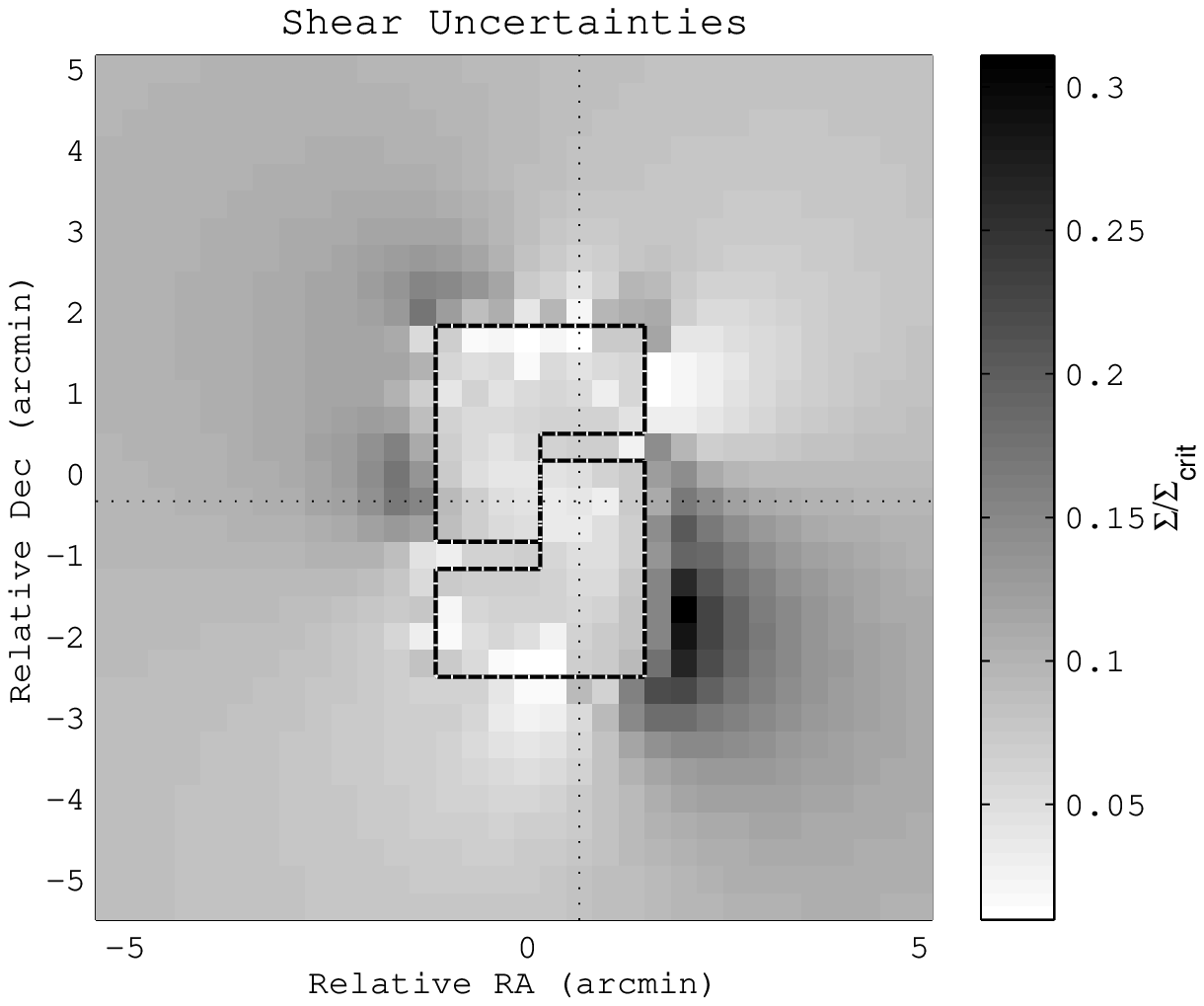,height=8cm, angle=90}}
\end{tabular}
}
\caption{
A smaller observed field.
(a) Mass distribution reconstructed using only shear information.
(b) Errors on the mass distribution reconstructed using only shear information.
The dotted lines show the lines along which
the cross sections are taken.\label{irr6_gg}}
\end{figure*}

\begin{figure}
\centerline{
\mbox{
\epsfig{file=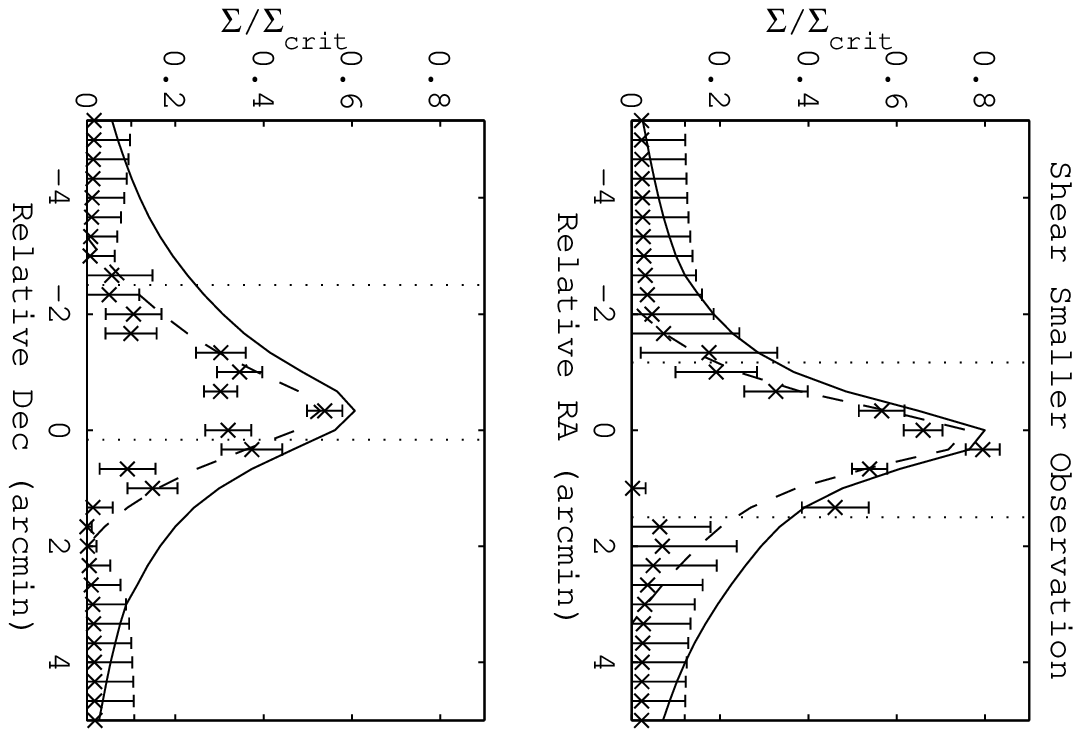,height=5.8cm, angle=90}
}
}
\caption{Two cross sections through the shear alone
reconstruction from only two WFPC2 type pointings.
The solid lines are slices through the original mass distribution; 
the dashed lines show the original mass distribution transformed by
the mass degeneracy transformation such that the minimum transformed
mass in the observed area is zero.
\label{irr6_slices}}
\end{figure}

We have explored the effects of using different numbers and
arrangements of WFPC2 type observations, and find that the results are
generally consistent with those presented above. The differences
occur when a much smaller area is observed. The reconstruction from
shear data alone tends to underestimate the mass due to the mass sheet
degeneracy effect. An example is shown in Fig. \ref{irr6_gg}, which
shows the reconstruction when only two WFPC2 type pointings are
used. Slices through this reconstruction are shown in Fig. \ref{irr6_slices}.
With this smaller observation, reconstruction from the shears alone
recovers $50$ per cent of the mass: $1.83\times 10^{14} h^{-1} M_{\odot}$ are
recovered in the (small) observed region compared with $3.62\times
10^{14} h^{-1} M_{\odot}$ for the original mass distribution. 
Note also that the mass concentration outside this observing field, to
the bottom left in Fig. \ref{origmass}, is still detected, despite the
fact that it is now outside the observation. 

The dashed lines in Fig.~\ref{irr6_slices} show the original mass
distribution transformed by the mass sheet degeneracy transformation
$\Sigma/\Sigma_{\rm{crit}} \rightarrow \lambda
\Sigma/\Sigma_{\rm{crit}} +(1-\lambda)$, where $\lambda$ is a free
parameter. 
We choose the value of $\lambda$ such that the minimum of the
transformed mass distribution in the observed field is zero
($\lambda=0.83$). The reconstructed points fit much better to this
line.
Qualitatively
 the mass reconstruction is correct relative to the mass
in the lowest mass pixel in the observed field. 

\begin{figure*}
\centerline{
\begin{tabular}{cccc}
(a)&
\mbox{\epsfig{file=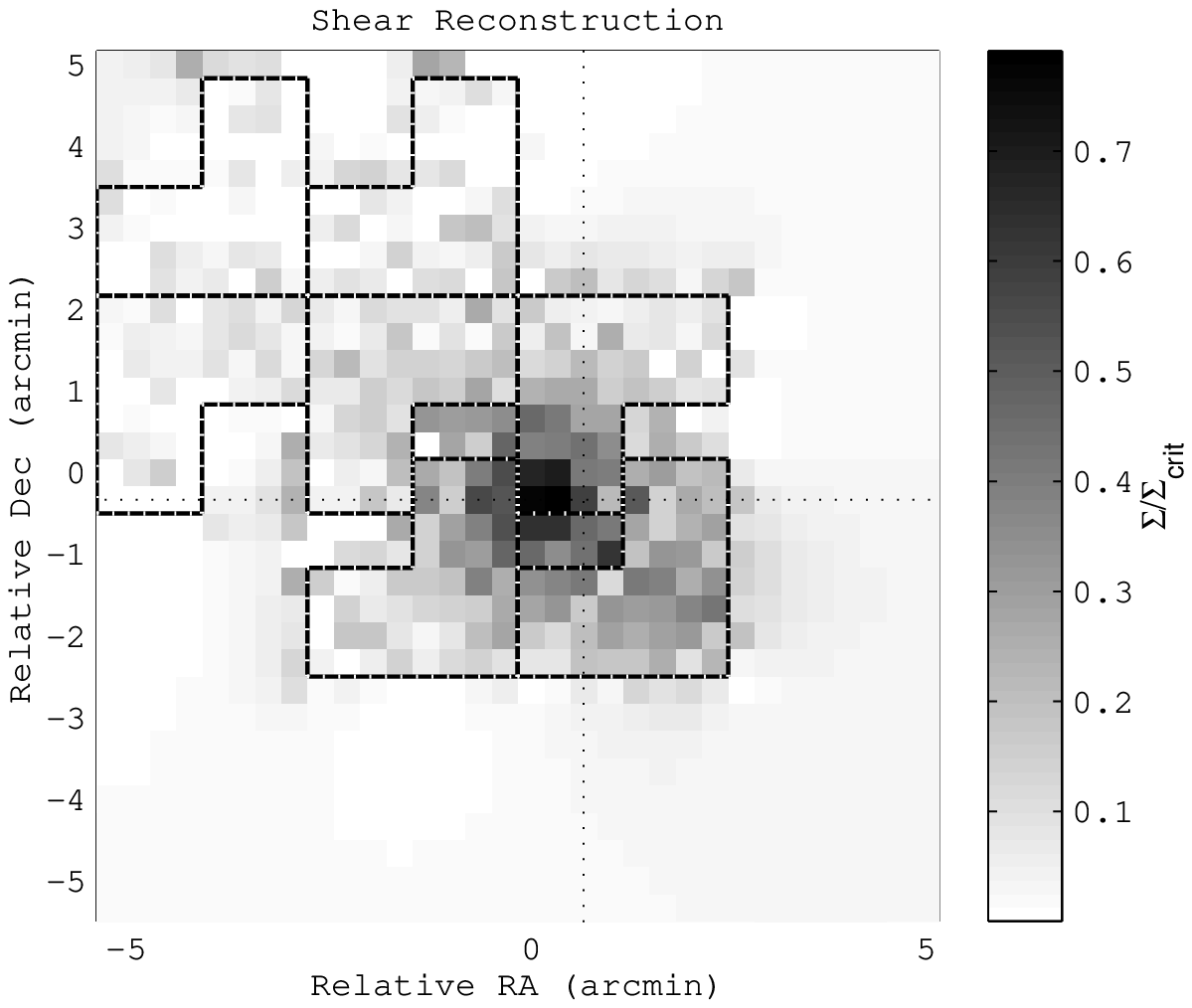,height=8cm, angle=90}}&
(b)&
\mbox{\epsfig{file=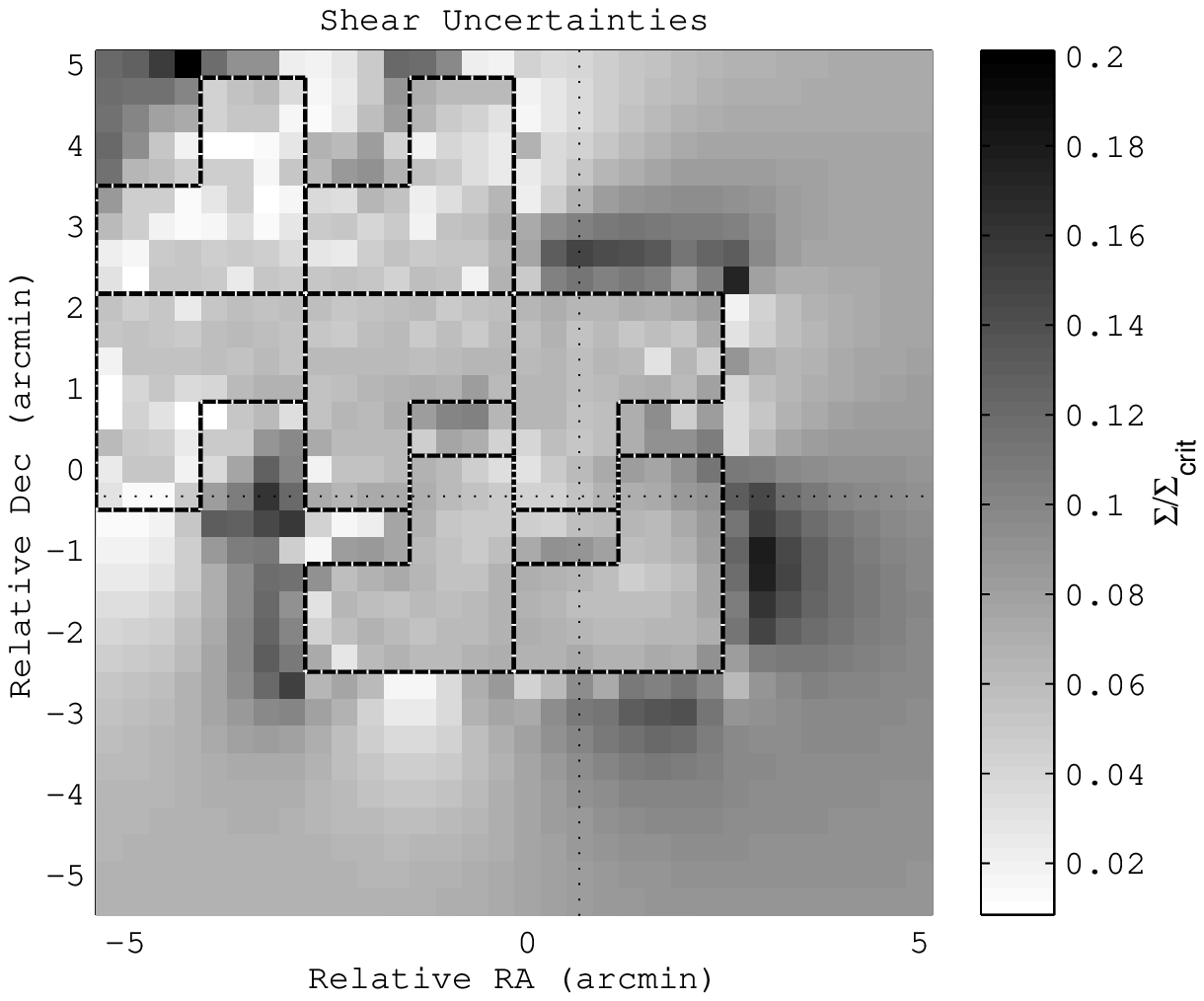,height=8cm, angle=90}}
\end{tabular}
}
\caption{
(a) Mass distribution reconstructed using only shear information.
(b) Errors on the mass distribution reconstructed using only shear information.
The dotted lines show the lines along which
the cross sections are taken.
\label{largerecon}}
\end{figure*}

\begin{figure}
\centerline{
\mbox{
\epsfig{file=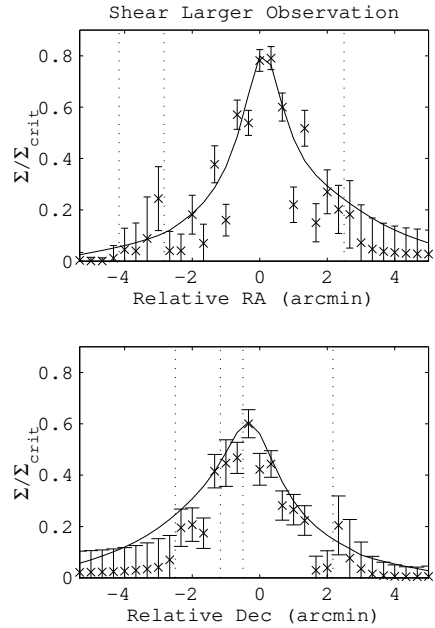,height=5.8cm, angle=90}
}
}
\caption{Two cross sections through the shear alone
reconstruction from only two WFPC2 type pointings.
\label{largeslices}}
\end{figure}

Fig.~\ref{largerecon} shows the result of extending the observed field beyond
that used in the main investigation of this paper. An additional 3
WFPC2 pointings are added to the top left of the cluster centre. Since
the mass density is so low at the edges of this observation (the
minimum value of the original mass distribution within the observed
region is zero)
we may hope that these extra observations 
will lift the mass degeneracy that occurs when shear data alone is
used. Slices through the reconstruction are shown in
Fig.~\ref{largeslices}.
As expected 
the reconstruction from the larger observed field is closer to the original
mass distribution. This is reflected in the fact that for this reconstruction
the total mass in the area covered by the 4 WFPC2 pointings is now
$76$ per cent of that for the original mass distribution, as
compared to $60$ per cent for the reconstruction from the 4 WFPC2
pointings.
It is interesting that on removing the extra WFPC2 pointing from the
middle left of the main 4 pointings $66$ per cent of the mass is
recovered; whereas if the top middle pointing is removed, $69$ per
cent is recovered; and if the top left pointing is removed $70$ per
cent is recovered, suggesting that a balance between extended and
continuous coverage has to be reached.
If the coverage pattern is extended to cover the whole $\sim$ 10
$\times$ 10 arc min area, $86$ per cent of the mass in the area
covered by the central 4 WFPC2  
pointings is recovered.

We also repeated this
investigation using a more centrally condensed mass distribution, with
$\theta_{\rm f}=5\,\,\theta_{\rm c}$ in Eq. \ref{sigmaeqn} instead of the more
conservative $\theta_{\rm f}=10\,\,\theta_{\rm c}$ used in this paper. In
this case reconstructing using shear data alone from 4 WFPC2 pointings
(same observing area as in Fig. \ref{images4}) 
recovers $77$
 per cent of the mass in the observed
region. When the observations are extended to the top left with 2 more
WFPC2 pointings $86$ per cent of the mass under the central 4 WFPC2
pointings is recovered. Finally, when the coverage pattern is extended to
cover the whole 10 $\times$ 10 arc min area, $100$ per cent of the mass
under the central 4 WFPC2 pointings is recovered.

\subsection{Coping with strong lensing regions}

\begin{figure*}
\centerline{
\begin{tabular}{cccc}
(a)&
\mbox{\epsfig{file=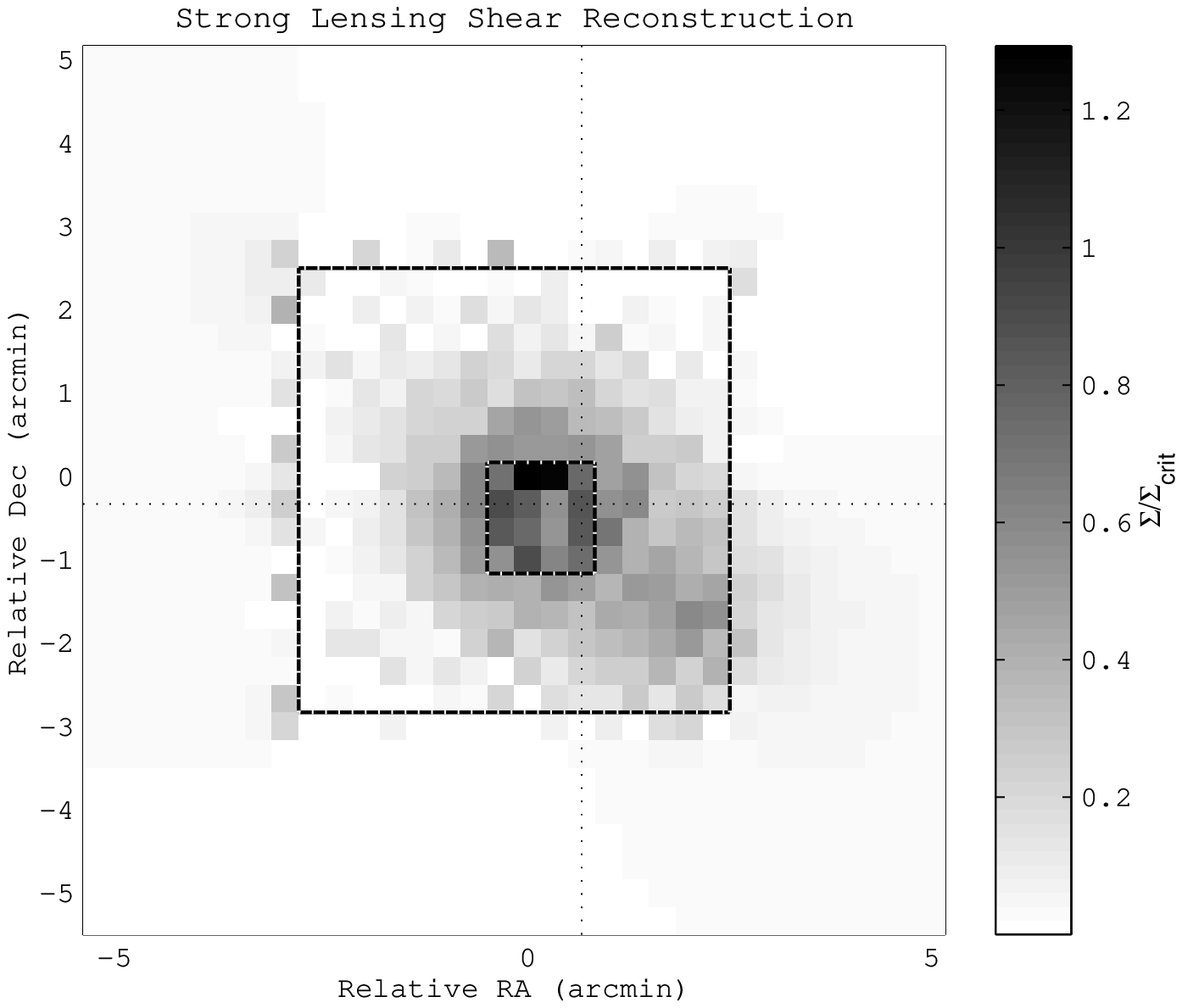,height=8cm, angle=90}}&
(b)&
\mbox{\epsfig{file=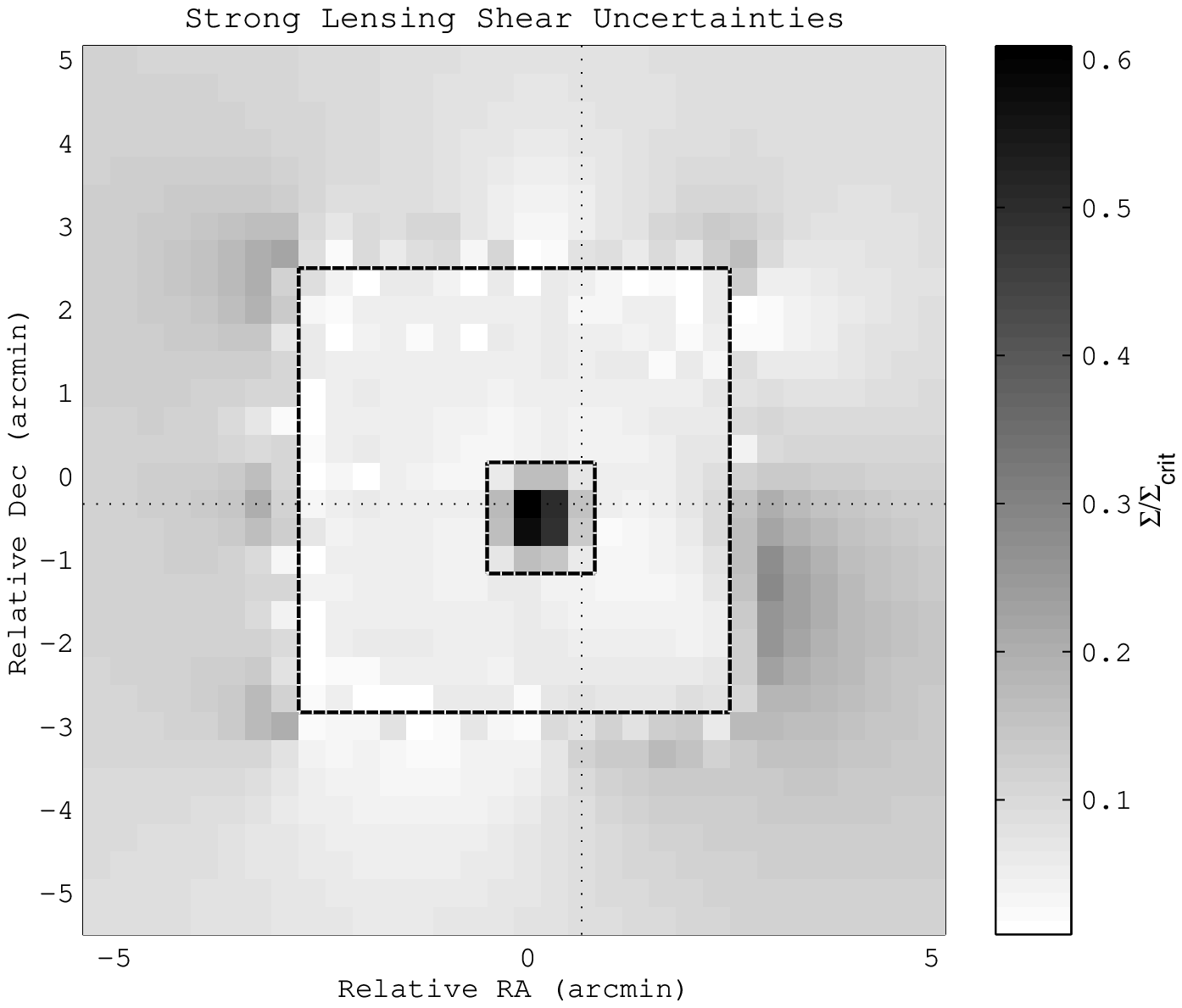,height=8cm, angle=90}}
\end{tabular}
}
\caption{
(a) Mass distribution reconstructed using only shear information.
(b) Errors on the mass distribution reconstructed using only shear information.
The dotted lines show the lines along which
the cross sections are taken.
\label{strongrecon}}
\end{figure*}

\begin{figure}
\centerline{
\mbox{
\epsfig{file=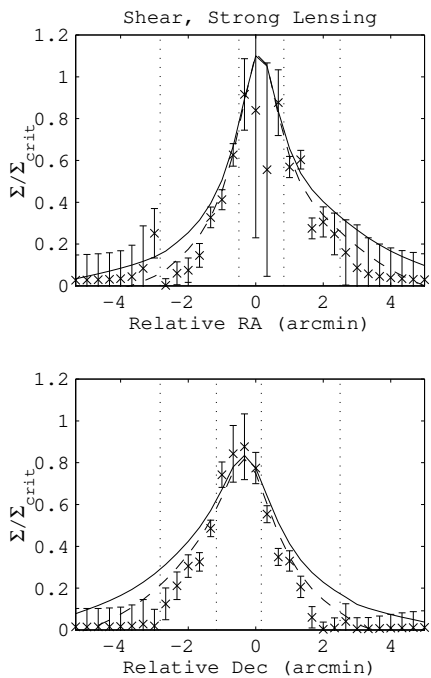,height=5.8cm, angle=90}
}
}
\caption{Two cross sections through the shear alone
reconstruction from only two WFPC2 type pointings.
The solid lines are slices through the original mass distribution; 
the dashed lines show the original mass distribution transformed by
the mass degeneracy transformation such that the minimum transformed
mass in the observed area is zero.
\label{strongslices}}
\end{figure}

In this subsection we reconstruct from an original mass distribution which is
identical to that shown in Fig.~\ref{origmass} but is multiplied by a
factor so that if plotted it would look identical to
Fig. \ref{origmass}, but the colour bar would range from $0$ to $1.1$.
Three potential problems with
using shear information from strong lensing regions 
($\Sigma/\Sigma_{\rm crit} >1$) are (1) ellipticities are not strictly an 
observable since we need to know whether or not a galaxy image has been 
inverted to calculate them; (2) the ellipticity map may no longer be
slowly varying but may vary on scales smaller than the pixel size;
(3) many galaxy ellipticity
estimation methods (e.g. Kaiser, Squires \& Broadhurst, 1995) are not
reliable when the shear becomes large (Erben et al. 2000). 
Therefore here we demonstrate a conservative
approach, in which we do not use data from strong lensing regions, but
do reconstruct through them. We do not use data from the central $16$ pixels.
The result of reconstructing from shear data alone 
is shown in Fig.~\ref{strongrecon}. 
The reconstruction follows
the original mass distribution well in the observed region, and the
mass distribution is reasonable given the large errors in the central 
$16$ pixels.
Notice that supercritical values have been reconstructed without any
particular problems.
For a more quantitative
comparison slices are shown in Fig.~\ref{strongslices}. 
Again, the dashed lines show the original mass distribution
transformed so that the minimum mass pixel in the observed area is
zero and the points follow this line better.
Clearly in practice one would want to use the strong lensing
information simultaneously with the weak lensing constraints
(AbdelSalam, Saha \& Williams 1998).


We have also carried out reconstructions in which the simulated data from the
central $16$ pixels \emph{is} used, together with the rest. The reconstruction
is much the same as before, except in the central $16$ pixels, where the result
is much more similar to the original, and the uncertainty map is uniform 
across the cluster. Thus despite the difficulties associated with data
from strong lensing regions the algorithm still produces reliable results.

\section{Conclusions}
\label{conclusions}

In investigating how well the BHLS maximum entropy method is capable
of deducing the matter distribution of typical massive galaxy clusters
at $z\approx0.4$ from typical shear data, from magnification data and
from both, we have demonstrated the following:
\begin{enumerate}
\item 
Shear data alone provide a useful estimate of the mass of
clusters: in simulations with realistic signal to noise and field of
view, 
the algorithm detects $50$ to $100$ per cent of the mass of the cluster
despite the mass degeneracy, depending on the size of the observation
and the profile of the cluster.
\item
The reconstruction method can
easily handle irregularly shaped and disjoint observing regions.
\item
Because shear information is a non-local function of the mass, small
gaps in observations can be bridged successfully when shear information
is used in the reconstruction.
\item
It is also successful at reconstructing mass distributions from
simulated magnification data alone.
\item
If the cluster has super-critical regions then the data from these
regions can be cut out and the reconstruction will bridge these
regions.
\end{enumerate}

\subsection*{ACKNOWLEDGMENTS}

SLB acknowledges financial support from a PPARC research studentship
and from the EEC TMR network `LENSNET'.
We thank Garret Cotter,
Ole M\"oller, Mike Jones, Jean-Paul Kneib, Alexandre Refregier, Andrew
Blain, Prasenjit Saha and Phil Marshall for helpful discussions.   

\bibliographystyle{/opt/TeX/tex/bib/mn}

\begin{thebibliography}{}

\bibitem{BNSS}
Bartelmann M., Narayan R., Seitz S., Schneider P., 1996, ApJ, 464, L115
\bibitem{bridlehls98} 
Bridle, S. L., Hobson, M. P., Lasenby, A. N., Saunders, R., 1998, \mnras, 299, 895
\bibitem{broadhurst}
Broadhurst T., Taylor A., Peacock J., 1995, ApJ, 438, 49
\bibitem{dyet98}
Dye, S., Taylor, A., 1998, \mnras, 300, L23
\bibitem{erbenvbms00}
Erben, T., Van Waerbeke, L., Bertin, E., Mellier, Y. \& Schneider,
P. 2000, A\&A submitted (astro-ph/0007021)
\bibitem{Gull90}
Gull S.F., Skilling J., 1990, The MEMSYS5 Users' Manual. Maximum Entropy Data Consultants Ltd, Royston.
\bibitem{Kaiser95} 
Kaiser N., 1995, ApJ, 439, L1
\bibitem{KS}
Kaiser N., Squires G., 1993, ApJ, 404, 441
\bibitem{mellier98}
Mellier, Y., 1998, to appear in ARAA Vol. 37, astro-ph/9812172
\bibitem{SS1}
Schneider P., Seitz C., 1995, A\&A, 294, 411
\bibitem{schneiderke00}
Schneider P., King L., Erben T., 2000, \aap, 353, 41
\bibitem{SS2}
Seitz C., Schneider, P., 1995, A\&A, 297, 287
\bibitem{SSB}
Seitz S., Schneider P., Bartelmann M., 1998, astro-ph/9803038
\bibitem{skilling}
Skilling J., 1989, in Skilling J., ed., Maximum Entropy and Bayesian
Methods. Kluwer, Dordrecht, p.~45
\bibitem{smail}
Smail, I., Ellis, R. S., Dressler, A., Couch, W. J., Oemler, A., Jr.,
Sharples, R. M., Butcher, H., 1997, \apj, 479, 70
\bibitem{squires96}
Squires G., Kaiser N., 1996, ApJ, 473, 65
\end{thebibliography}

\bsp % ``This paper has been produced using the ...''
\label{lastpage}

\end{document}